\documentclass[reqno,pre,showpacs,superscriptaddress,11 pt]{revtex4}

\usepackage{epsfig}
\usepackage{graphicx,color}
\usepackage{verbatim}
\usepackage{amssymb,latexsym,mathrsfs,footnote}
\usepackage[centertags]{amsmath}
\usepackage{amsfonts,amsmath}
\usepackage{amsthm}
\usepackage{newlfont}
\usepackage{subfigure,changes}

\newcommand{\opunit}{\text{1}\kern-0.22em\text{l}}

\newcommand{\id}{\textrm{d}}

\newcommand{\beq}{\begin{equation}}
\newcommand{\eeq}{\end{equation}}

\def\bea{\begin{eqnarray}}
\def\eea{\end{eqnarray}}
\def\ba{\begin{array}}
\def\ea{\end{array}}
\def\be{\begin{equation}}
\def\ee{\end{equation}}

\def\la{\langle}
\def\ra{\rangle}

\begin{document}

\title{Short-time Behaviour  and Criticality of Driven Lattice Gases}

\author{Urna Basu}
\affiliation{SISSA -- International School for Advanced Studies and INFN, via Bonomea 265, I-34136 Trieste}
\author{Valerio Volpati}
\affiliation{SISSA -- International School for Advanced Studies and INFN, via Bonomea 265, I-34136 Trieste}
\author{Sergio Caracciolo}
\affiliation{Universit\`a degli Studi di Milano -- Dip.~di Fisica and INFN, via Celoria 16, I-20133 Milano}
\author{Andrea Gambassi}
\affiliation{SISSA -- International School for Advanced Studies and INFN, via Bonomea 265, I-34136 Trieste}

\begin{abstract}
The nonequilibrium short-time critical behaviors of driven and undriven lattice gases
are investigated via Monte Carlo simulations in two spatial dimensions starting from a fully disordered
initial configuration. In particular, we study the time evolution of
suitably defined order parameters, which account for the strong anisotropy
introduced by the homogeneous drive.
We demonstrate that, at short times, the dynamics of all these models is unexpectedly described by an effective continuum theory in which transverse fluctuations, i.e., fluctuations averaged along the drive, are Gaussian, irrespective of this being actually the case in the stationary state. Strong numerical evidence is provided, in remarkable agreement with that theory, both for the driven and undriven lattice gases, which therefore turn out to display the same short-time dynamics. 

\end{abstract}

\pacs{64.60.Ht, 05.70.Ln, 05.10.Ln, 05.50.+q }

\maketitle

Understanding the emergence of collective behaviours in statistical systems out of equilibrium is one of the major challenges of modern physics. Within a bottom-up approach, simplified lattice models were proposed in the past in order to capture specific aspects of nonequilibrium physics, e.g., the presence of external forces which induce steady currents in a system. 
The driven lattice gas (DLG) is perhaps the simplest among them: it generalizes the equilibrium 
lattice gas (LG) \cite{LeeYang} to account
for the presence of an external driving field $E$ which acts along one lattice direction, biasing the particle jump rates. 
The DLG was introduced in 1983 \cite{KLS} and it has since become the paradigm for driven diffusive systems \cite{DDSbook,Zia25}. At half filling, for any value of $E$, its stationary state shows a nonequilibrium continuous transition from a disordered to an ordered state at a critical temperature $T_c(E)$. In the ordered ``phase'' particles and holes are separated by an interface parallel to the direction of $E$ \cite{KLS,DDSbook,Zia25}.  The drive introduces also a strong spatial anisotropy and the observed transition differs from that of the Ising universality class $(E=0)$.

In spite of the apparent simplicity of the model, the critical behaviour of the DLG has been 
a matter of debate for the past three decades. Early field-theoretical studies \cite{JS} proposed that the evolution of DLG is effectively described by a mesoscopic Langevin equation with spatial anisotropy and a finite particle current near the critical point. Within this theory (referred to as JSLC, from the authors' names), the critical behaviour of the transverse fluctuations turn out to be non-interacting (Gaussian). Earlier numerical studies  partially supported this prediction \cite{Valles,Wang,LeungFS}.

However, some important discrepancies \cite{Garrido1994, Marro1996} cast doubts on
this conclusion and in fact an alternative description was introduced \cite{Garrido1998} according to which in the DLG with infinitely strong drive (IDLG) $E=\infty$, only anisotropy and not current is the relevant nonequilibrium effect. 
As a consequence, IDLG is described by the same effective model as the randomly driven lattice gas (RDLG), i.e.,  a DLG with $E$ randomly changing sign. This proposal was contradicted by subsequent studies \cite{Leung3,Schmittmann00} and by a more careful finite-size scaling analysis \cite{Caracciolo1,Caracciolo2}. However, this debate continued with the numerical study in Ref.~\cite{Garrido2001}, which found the same finite-size scaling functions in the stationary state for both the IDLG and the RDLG.

The numerical study of the DLG is affected by the particularly severe critical slowing down, typical of systems with a local conservation law. Moreover, extracting information from numerical data is difficult because of the peculiarities of finite-size scaling in the presence of strong anisotropy~\cite{CGGP}.  A way to bypass this issue  
is to study the short-time dynamical relaxation towards the stationary state following a quench to the critical point \cite{JSS,Zheng1998}. For the DLG, this was done in Ref.~\cite{AS2002} (see also \cite{replyAS}) and the resulting  critical exponents turned out to be the same as those of the RDLG, even for finite $E$. However, this conclusion has recently been questioned again in Ref.~\cite{Tauber},  which revisits the short-time critical dynamics and ageing in the IDLG and shows agreement with the JSLC theory.

In this Letter, we show that in these lattice gases, the short-time behaviour of `transverse' observables, i.e., of quantities which have been spatially averaged along the direction of $E$, is dictated by an effective Gaussian theory, rather independently of the actual microscopic dynamics. 
To this end we perform extensive Monte Carlo simulations of the IDLG, RDLG and LG, showing that the time evolution of suitably defined transverse order parameters agrees excellently with the prediction of the Gaussian theory with one exception in the LG. 
We also show that, in the stationary state and in the thermodynamic limit, the IDLG still displays this Gaussian behaviour, while the RDLG \emph{does not}.   
Accordingly, the critical behaviour of the driven lattice gases cannot actually be distinguished from studying the short-time dynamics of transverse observables. Our findings actually encompass and reconcile a number of apparently contradictory statements, hopefully settling a twenty years long debate on a set of paradigmatic models.

{\bf Models:} The LG is defined on a $d$-dimensional hypercubic lattice where each site $i$ can be either occupied by one particle or empty, with occupation number $n_i = 1$ or $0$, respectively. The particles jump randomly to empty nearest-neighbour sites with rates $w(\Delta\mathcal H) = \min \{1, e^{-\beta \Delta \mathcal H} \}$ where  $\Delta \mathcal H$ is the change in the nearest-neighbour attractive Hamiltonian $\mathcal{H} =-4  \sum_{\{ i,j \}} n_i n_j$ due to the proposed jump and $\beta = 1/T$ is the inverse temperature.
At half filling, in the thermodynamic limit and at the critical temperature  $T_c = 2/\log(1 + \sqrt{2})$,   the LG undergoes a continuous phase transition belonging to the Ising universality class~\cite{Mussardo}.

The DLG is obtained from the LG by adding a constant field $E$ 
along one lattice axis, which biases the jump rates as $w(\Delta\mathcal H + E l)$ where 
$l=1,-1,0$ for jumps occurring along, opposite, or transverse to the field. This dynamics leads to a nonequilibrium stationary state when the boundary conditions are periodic along the field direction, as detailed balance is broken. 
At half filling and in the thermodynamic limit,  the DLG shows a phase transition at the critical temperature $T_c(E)$ below which particles condense in a single strip with interfaces parallel to the direction of $E$ \cite{DDSbook}. For $E \to \infty$ a particle jump along (opposite to) $E$ is always accepted (rejected). This case is referred to as the IDLG and here we focus on it. 

The randomly driven lattice gas (RDLG) \cite{rdlg1, rdlg2} is a variant of the DLG in which the field $E$ changes sign randomly at each attempted move;  for simplicity, we consider below $E=\pm\infty.$ Similarly to the DLG, the RDLG at half filling undergoes a continuous transition to a phase-separated configuration. 

We perform Monte Carlo simulations on a rectangular  periodic lattice of size $L_{\parallel} \times L_\perp$ where $\parallel$ and $\perp$ denote the directions parallel and orthogonal to the drive, respectively, with ``volume'' $V = L_\parallel L_\perp$ ; each Monte Carlo step consists of $V$ attempted jumps and it sets 
the unit of time. The evolution is studied at the critical temperature $T_c$ starting from a fully disordered configuration, which is equivalent to a quench from $T = \infty$.

The onset of order in the DLG is typically characterized via the so-called anisotropic order parameter
\be
m = \left \la \; | \mu | \; \right\ra/V,
\label{eq:def-m}
\ee
which is the statistical average $\la \cdots\ra$ of the first non-zero transverse mode $\mu = \tilde \sigma(0,2 \pi/L_\perp)$, where 
 \bea
\tilde{\sigma}(k_{\parallel}, k_{\perp}) = \sum_{x=0}^{L_{\parallel}-1} \sum_{y=0}^{L_{\perp}-1} e^{i(k_{\parallel} x + k_{\perp} y )} \sigma_{xy}. 
\label{eq:sigma-FT}
\eea
 Here $\sigma_{xy} = 2 n_{xy}-1 $ is the spin variable associated with each site $(x,y)$ and $(k_{\parallel}, k_{\perp}) = \left(2 \pi n_{\parallel}/ L_\parallel, 2 \pi n_\perp/L_\perp \right)$ denote the parallel and the transverse wavevectors  with integers $0 \le n_{\parallel,\perp} \le  L_{\parallel,\perp} -1.$

An alternative order parameter was introduced in Ref.~\cite{AS2002} to measure the average absolute value of the magnetization of the lines parallel to $E$, which can also be expressed as a sum of transverse modes,
\be
O = \frac 1 V \sum_{y=0}^{L_\perp-1} \bigg \la \bigg| \sum_{x=0}^{L_{||}-1} \sigma_{xy} \bigg| \bigg \ra = \frac 1 V \bigg \la \bigg| \sum_{n_\perp=1}^{L_\perp-1} \tilde{\sigma}\left(0, \frac {2 \pi n_\perp}{L_\perp}\right) \bigg| \bigg \ra.
\label{eq:def-O}
\ee

In contrast to the LG, both  DLG and  RDLG show 
strong anisotropy in space and the finite-size scaling analysis has to be done at a fixed aspect ratio $S_\Delta = {L_{\parallel}}/{L_\perp^{1+\Delta}}.$ 
The strength of the anisotropy is characterized by 
$\Delta = \nu_\parallel/\nu -1$  where $\nu$ and $\nu_\|$ are the critical exponents of the correlation length along the transverse  and parallel directions, respectively. As a result, all the critical exponents (see Table \ref{tab:d2}) depend on the direction \cite{DDSbook}.

{\bf Gaussian theory:} The mesoscopic description of the DLG \cite{JS} is based on a Langevin equation for the coarse-grained local particle density $\rho(x,t)$. Near criticality, 
the evolution of the spin density $\phi(x,t) = 2 \rho(x,t) -1$ reads, 
\be
\begin{split}
\partial_t \phi =& \alpha [(\tau - \nabla_\perp^2)\nabla_\perp^2 \phi + \tau_\parallel \nabla_\parallel^2 \phi + {\cal E}\nabla_\parallel \phi^2  ] \cr
& + u \nabla_\perp^2 \phi^3 - \nabla_\perp \cdot \xi .
\end{split}
\label{eq:L-DLG}
\ee
Here $\tau$ measures the distance from critical point, ${\cal E}$ represents the coarse-grained $E$, 
and $\xi$ is a white noise with $\la \xi_i(x,t) \xi_j(x',t')\ra \propto \delta_{ij}\delta^d(x-x')\delta(t-t')$, while $\tau_\|$, $\alpha$, and $u$ are inconsequential positive constants. 
As the only relevant interaction in Eq.~\eqref{eq:L-DLG} is ${\cal E}\nabla_\parallel \phi^2$, the order parameter $\phi$ at vanishing parallel wavevector $k_\parallel= 0$ behaves as in a non-interacting theory \cite{JS} and therefore transverse fluctuations are Gaussian \cite{Caracciolo1}. 
It is then natural to investigate the consequences of this strong prediction on the short-time behaviour of transverse modes. 

 \begin{table}
\begin{tabular}{c|c|c|c}
\hline
Exponent & ~~~JSLC~~~ & ~~~RDLG~~~ & ~~LG~~ \cr
\hline\hline
$\Delta$ & 2 & 0.992 & 0 \cr
$\beta$ & 1/2 & 0.315 & 1/8 \cr
$\nu$ & 1/2 & 0.626 & 1 \cr
$\eta$ & 0 & 0.016 & 1/4 \cr
$z$ & 4 & 3.984 & 15/4 \cr
\hline\hline
\end{tabular}
  \caption{Critical exponents in $d=2$ for the JSLC \cite{JS}, RDLG \cite{rdlg1}, and LG \cite{Mussardo}. 
The values listed for the JSLC and the RDLG refer to the transverse exponents; the values for the JSLC and LG are exact, while those for the RDLG are obtained approximately from a series expansion.}
  \label{tab:d2} 
 \end{table}

The equation of motion for the amplitude $\tilde{\sigma}_k \equiv \tilde\sigma(0,k)$ of any transverse mode follows from Eq.~\eqref{eq:L-DLG}
\beq
\frac{\id}{\id t} \tilde{\sigma}_k(t)= - \gamma_k \tilde{\sigma}_k(t) + i\hat k\, \eta_k(t), \label{eq:g_evol}
\eeq
where  $\hat k = 2 \sin (k/2) $ and $\gamma_k= \lambda (\tau + \hat k^2) \hat k^2$.
$\lambda$ is a coarse-grained diffusion constant and $\eta$ is the white noise with $\la \eta_k(t)\eta_{k^\prime}(t^\prime)\ra  = 2 \lambda Z V \delta (k+k^\prime)\delta (t-t^\prime)$ where $Z$ is a normalization factor. 

We are interested in the case of a quench to the critical point from a disordered configuration, i.e., $\tilde{\sigma}_k(t=0)=0$ for all $k$.
For this choice, Eq.~\eqref{eq:g_evol} implies a Gaussian probability distribution for 
$\tilde{\sigma}_k(t)$ 
\be
P[\tilde{\sigma}_k(t),\tilde{\sigma}^{*}_{k}(t)]~  
\propto ~ \exp {\left\{- \frac{|\tilde{\sigma}_k(t)|^2}{V \tilde{G}_{\perp}(t,k)} \right\}}
\label{eq:Psigma}
\ee
at any time $t$, where $\tilde{G}_{\perp}(t,k) = \la |\tilde \sigma_k(t) |^2 \ra / V$ is the transverse propagator,  
which is easily determined from Eq.~\eqref{eq:g_evol}; at the critical point $\tau=0,$ $\tilde G_\perp(t,k) = Z(1- e^{-2 \lambda t \hat k^4})/\hat k^2.$

The order parameter $m$ can now be calculated from $P$ by taking the average according to Eqs.~\eqref{eq:def-m} and  \eqref{eq:Psigma}
\be
m(t) =  \sqrt{\frac \pi {4V} G_\perp \left(t,\frac{2\pi}{L_\perp} \right)}  \sim \sqrt{\frac t{L_{\parallel} L_{\perp}^3}}, \label{eq:mt}
\ee
where the last expression indicates the behaviour at short times $t \ll L_\perp^4$ on  large lattices 
$L_\perp \gg \pi$. 

The evolution of $O$ in Eq.~\eqref{eq:def-O} can be determined analogously, by noting that the sum of a set of Gaussian-distributed variables is also Gaussian. Accordingly, 
\be
O(t) = \sqrt{\frac \pi 4 \frac 1 V \sum_{n_\perp=1}^{L_\perp -1} \tilde G_\perp \left(t,\frac{2 \pi n_\perp}{L_\perp} \right)}. 
\label{eq:O-sum}
\ee
For large $L_\perp$ the sum over $n_\perp$ turns into an integral, which yields $Z L_\perp (2\lambda t)^{1/4}/\pi$ and 
\be
O(t) \sim  t^{1/8}/L_{\parallel}^{1/2} 
\label{eq:Ot}
\ee 
to the leading order for $t\ll L_\perp^4$.
The time dependences predicted in Eqs.~\eqref{eq:mt} and \eqref{eq:Ot} were indeed observed in previous numerical studies of short-time dynamics~\cite{AS2002,Tauber}.

At criticality, a phenomenological scaling analysis \cite{AS2002,Tauber} for $m$ yields
\bea
m (t, L_{\parallel} ; S_\Delta) = L_{\parallel}^{-\beta/[\nu(1+\Delta)]} \tilde f_m (t/L_{\parallel}^{z/(1+\Delta)
} ; S_\Delta).
\label{eq:scalingass}
\eea
Remarkably, the prediction in \eqref{eq:mt} of the Gaussian theory is  compatible with this scaling, \emph{independently} of the specific set of values of the critical exponents used, namely those of JSLC, RDLG, and LG in Tab.~\ref{tab:d2}. %
This compatibility is a direct consequence of the hyperscaling relation $ d + \Delta - 2 + \eta = 2 \beta / \nu$ \cite{DDSbook} valid for all three sets, and therefore, in the presence of Gaussian fluctuations, Eq.~\eqref{eq:scalingass} at short times is not capable of distinguishing between the various universality classes, contrary to what was assumed in Ref.~\cite{Tauber}. The same scaling form for $O,$ with a different scaling function $\tilde f_O,$ is compatible with the prediction of Gaussian theory \eqref{eq:Ot} only if $\eta=0,$ which is exactly (approximately) true for JSLC (RDLG). Accordingly, in contrast with the assumption in Ref.~\cite{AS2002},  $O$ is also not able to distinguish between these two universality classes. For LG $\eta \ne 0$  thus \eqref{eq:Ot} does not hold; however, assuming $O \sim L_{||}^{-1/2}$ \cite{AS2002} we get $O(t) \sim t^{1/10}$ from the scaling analysis \cite{SM}.   

\begin{figure}[h]
 \centering
 \includegraphics[width=12 cm]{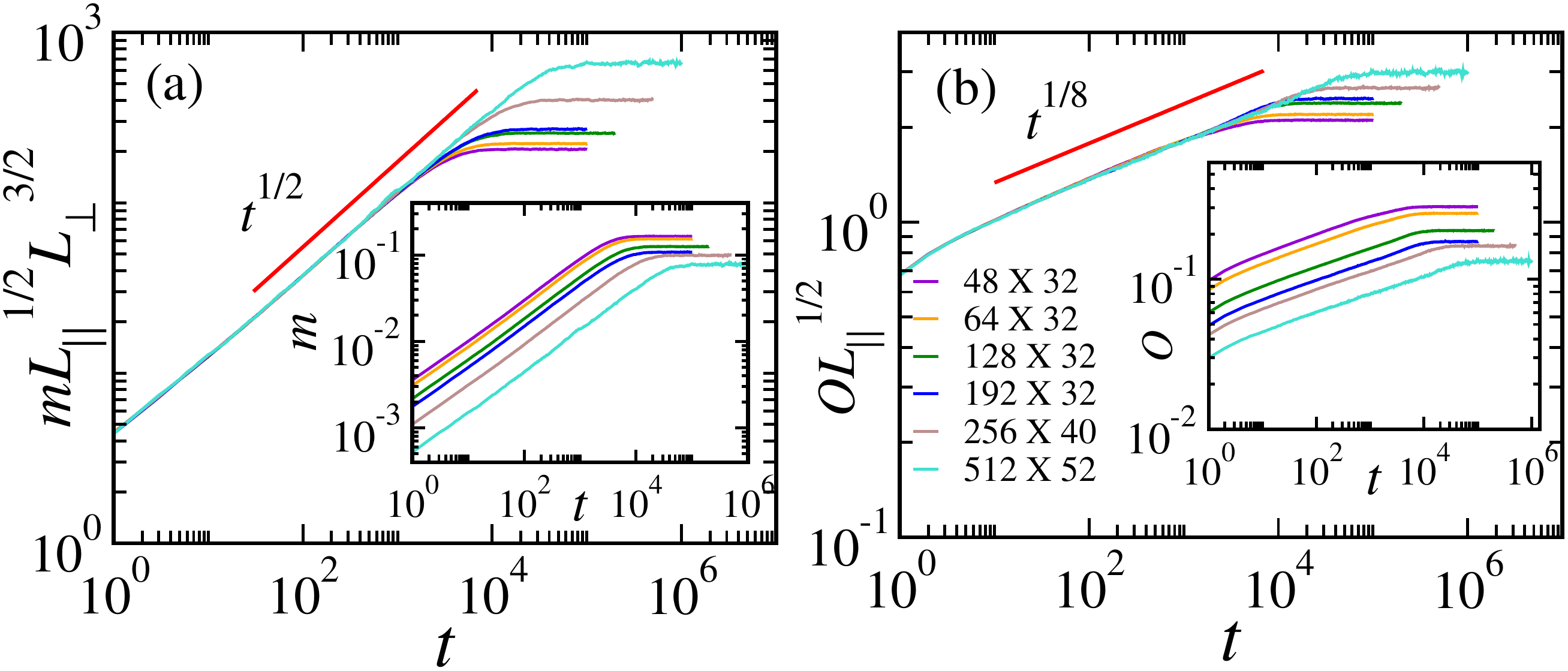} 
 \caption{(Color online) Short-time behaviour of the IDLG: Scaling plot of (a) $m(t)$ and (b) $O(t)$ [see Eqs.~\eqref{eq:def-m} and \eqref{eq:def-O}] compared with the predictions in Eqs.~\eqref{eq:mt} and \eqref{eq:Ot}, respectively for different lattice sizes $L_\|\times L_\perp$. In both panels, the insets show the corresponding unscaled data. 
}
 \label{fig:idlg}
\end{figure}
%

A useful indicator of possible deviations from Gaussian behaviour is the Binder cumulant, which, for conserved systems, can be suitably defined as \cite{Caracciolo2}
\bea
g = 2- \la |\mu|^4 \ra/\la |\mu|^2 \ra^2 
\label{eq:binder} 
\eea
where $\mu$ is the  lowest transverse mode. 
Indeed, a non-zero value of $g$ signals non-Gaussian transverse fluctuations. 
The JSLC theory in Eq.~\eqref{eq:L-DLG} predicts that the stationary value of $g$ at the critical point vanishes upon increasing the system size, as it was verified in Ref.~\cite{Caracciolo2} for the IDLG.

Below we test the validity of the theoretical predictions discussed above via Monte Carlo simulations in $d=2$.

{\bf IDLG:} The evolution of the IDLG is studied at the critical temperature $T_c=3.20$ \cite{KLS}, starting from a fully disordered configuration where both $m$ and $O$ vanish. Figure \ref{fig:idlg}(a) shows $m$ as a function of time $t$ for various geometries  with no fixed $\Delta$.
The data follow the prediction \eqref{eq:mt} of the Gaussian theory and an excellent collapse is indeed obtained by rescaling the raw data for $m$ in the inset by $L_\parallel^{1/2} L_\perp^{3/2}$, with a growth $\sim t^{1/2}$.
Figure \ref{fig:idlg}(b) shows the evolution of   $O$, confirming Eq.~\eqref{eq:Ot}; the scaling collapse is obtained by plotting $O(t)L_{\parallel}^{1/2}$  which grows as $t^{1/8}.$

\begin{figure}[thb]
 \centering
 \includegraphics[width=12 cm]{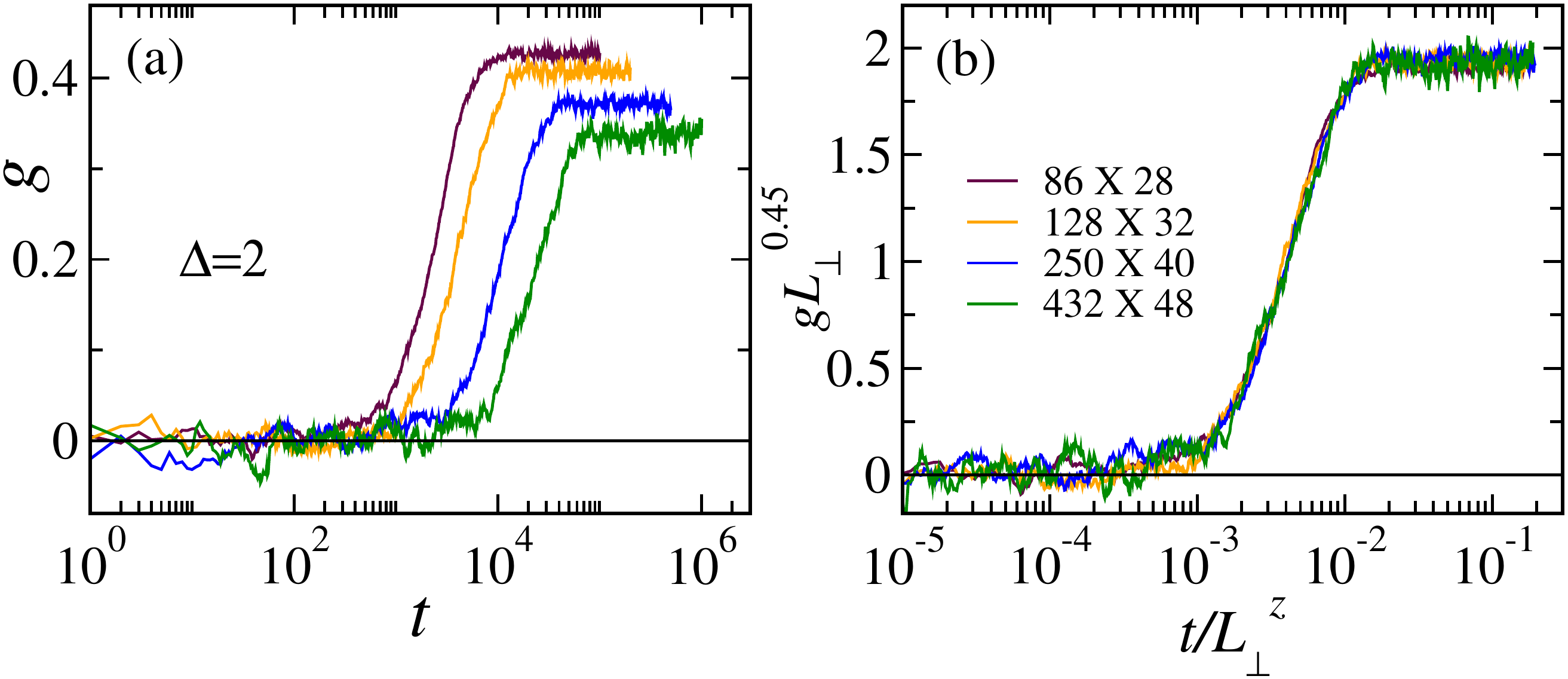} 
\caption{(Color online) Binder cumulant $g$ [see Eq.~\eqref{eq:binder}] of the IDLG for the lattice sizes $L_\|\times L_\perp$ 
with fixed  $S_\Delta = 2^{-8}$ and $\Delta=2$. The raw data of panel (a) collapse on a single curve in panel (b) when $g L_\perp^{0.45}$ is plotted vs.~$t/L_\perp^z$, with $z=4.$
 }
 \label{fig:binder_idlg}
\end{figure}
%

The evolution of the Binder cumulant $g$ defined in Eq.~\eqref{eq:binder} 
is shown in Fig.~\ref{fig:binder_idlg}(a) for various geometries corresponding to a fixed $S_\Delta = 2^{-8}$ with $\Delta=2$. Consistently with a Gaussian behaviour at short times, 
$g$ 
is vanishingly small 
up to a certain time scaling as $\sim L_\perp^z$ for large $L_\perp$, eventually reaching a stationary value which decreases upon increasing $L_\perp$.
In fact, it was shown in Ref.~\cite{Caracciolo2} that the stationary value of $g \sim L_\perp^{-0.45(15)}$. This is verified in Fig.~\ref{fig:binder_idlg}(b) which shows the perfect scaling collapse of the same data as panel (a) multiplied by $L_\perp^{0.45}$ and plotted as a function of $t/L_\perp^z$.  

The numerical evidences reported in Figs.~\ref{fig:idlg} and \ref{fig:binder_idlg} demonstrate that, in the IDLG, the short-time behaviour of transverse modes is Gaussian for any finite system, with arbitrary geometry, whereas this is not the case in the stationary state on a finite lattice; however, in the thermodynamic limit, the Gaussian behaviour predicted by JSLC is recovered because $g$ vanishes. 

%
\begin{figure}[hb]
 \centering
 \includegraphics[width=12 cm]{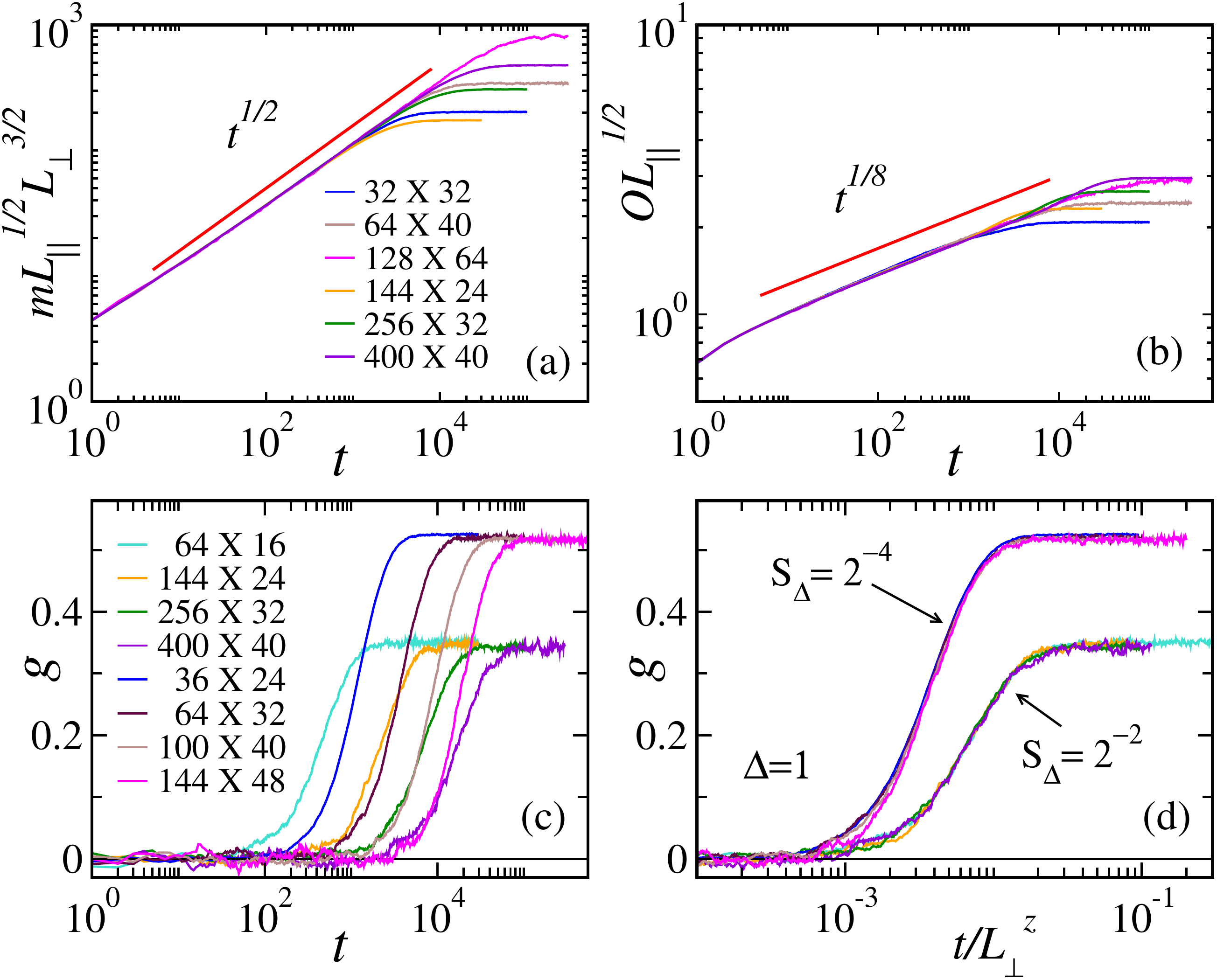} 
 \caption{(Color online) 
Evolution of the RDLG: Scaling plot of (a) $m(t)$ and (b) $O(t)$ [see Eqs.~\eqref{eq:def-m} and \eqref{eq:def-O}]  compared with the predictions in Eqs.~\eqref{eq:mt} and \eqref{eq:Ot}, respectively for different lattice sizes $L_\|\times L_\perp$.  
(c) Evolution of the Binder cumulant $g$ in various systems with $\Delta =1$ and two different values of $S_\Delta =2^{-4}$ and $2^{-2}$. (d) Same data as in (c) but plotted as a function of $t/L_{\perp}^z$, with $z \simeq 4$, according to the RDLG theory. }
 \label{fig:rdlg}
\end{figure}
%

{\bf RDLG:}  The evolution of the RDLG is investigated after a quench to the critical point  $T_c = 3.15$. Figures \ref{fig:rdlg}(a) and \ref{fig:rdlg}(b) show that the numerical data for $m$ and $O$ follow  Eqs.~\eqref{eq:mt} and \eqref{eq:Ot}, 
respectively, as predicted by the Gaussian theory and irrespective of any specific $S_\Delta$.
In this respect, RDLG and IDLG are indistinguishable.

This Gaussian behaviour at short times  is also evident from
the fact that, independently of the lattice sizes and similarly to the IDLG, $g$ is initially vanishingly small, as shown in Fig.~\ref{fig:rdlg}(c) and then it increases 
towards its stationary value.  Figure \ref{fig:rdlg}(d) shows the same data as in panel (c) but plotted as a function of $t/L_\perp^z$ for $\Delta \simeq 1$ and two values of $S_\Delta$.
In contrast to the IDLG, the stationary value of $g,$ for a fixed $S_\Delta,$ does {\it not} change upon increasing the system size, confirming that IDLG and RDLG display different stationary critical behaviours.

{\bf LG:} Surprisingly, the IDLG and RDLG show a similar Gaussian behaviour at short times, 
in spite of their different microscopic dynamics. This might be due to the presence of particle conservation in their dynamics. 
It is then instructive to study critical quenches of the LG, starting from the same initial conditions as in the IDLG and RDLG.  LG being isotropic ($\Delta=0$), the  transverse direction is chosen arbitrarily. 
%
\begin{figure}[th]
 \centering
 \includegraphics[width=12 cm]{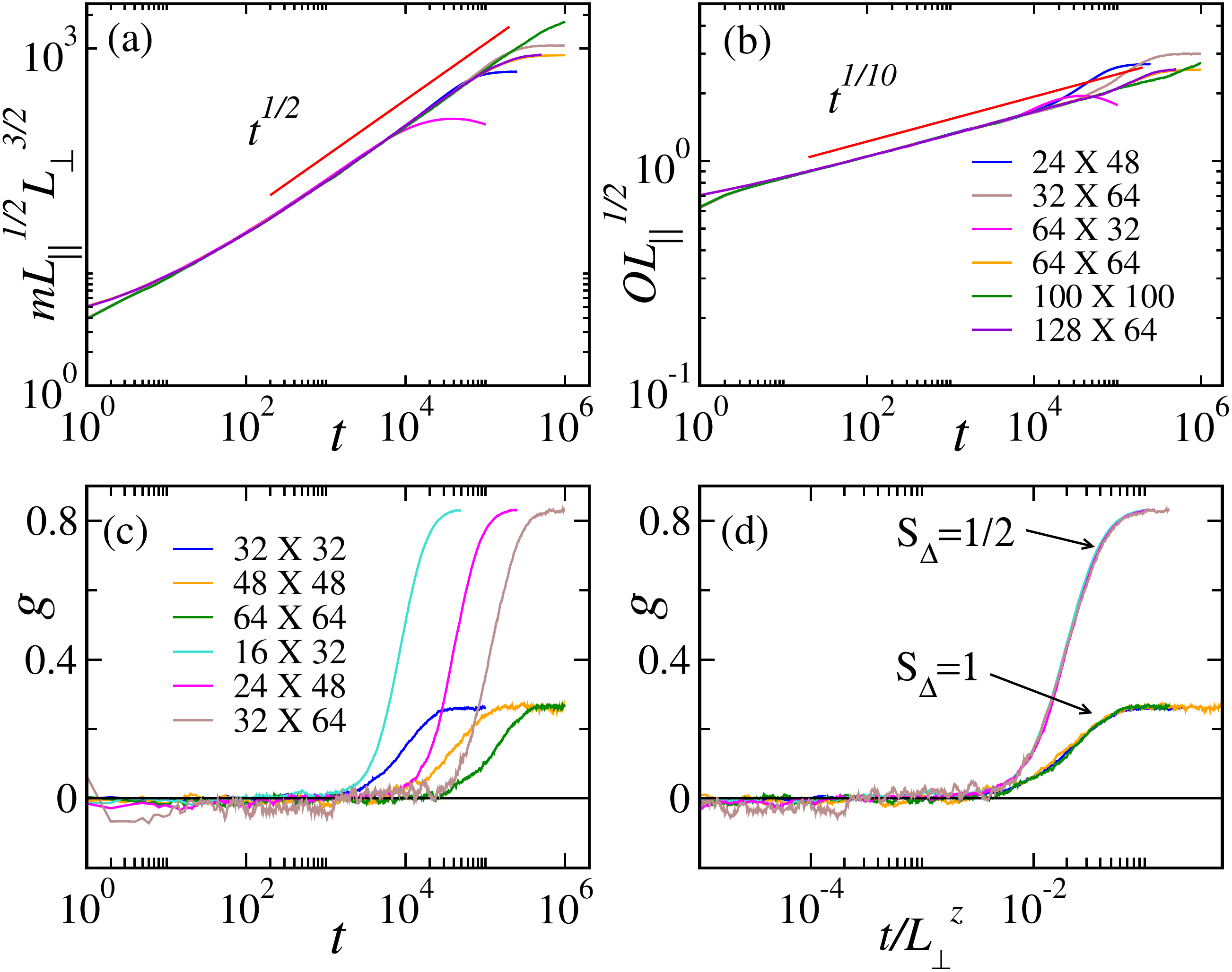} 
 \caption{(Color online) 
Evolution of the LG: Scaling plot of (a) $m(t)$ and (b) $O(t)$ [see Eqs.~\eqref{eq:def-m} and \eqref{eq:def-O}]  compared with the predictions in Eq.~\eqref{eq:mt} and with a growth $\sim t^{1/10}$, respectively, for different lattice sizes $L_\|\times L_\perp$.  
(c) Evolution of the Binder cumulant $g$ in various systems with $\Delta = 0$ and two different values of $S_\Delta =1/2$ and $1$. (d) Same data as in (c) but plotted as a function of $t/L_{\perp}^z$, with $z = 3.75$. The resulting scaling curves depend on $S_\Delta$.
}
 \label{fig:ising}
\end{figure}
%
%
Figure \ref{fig:ising}(a) shows $m(t)$ and its scaling according to the prediction $\sim t^{1/2}$ of the Gaussian theory 
for various geometries, showing an excellent collapse. 
The behaviour of $O,$ however, is different from Eq.~\eqref{eq:Ot}, as already pointed out; it shows a $t^{1/10}$ growth, while the finite size scaling $\sim L_{\parallel}^{-1/2}$ is still the same as in the other cases. 
Figure~\ref{fig:ising}(c) shows that, similar to IDLG and RDLG, the Binder cumulant $g$ stays close to zero up to a certain time which scales $\sim L_{\perp}^z$ with $z=3.75$ (see Table \ref{tab:d2}), as can be inferred from the perfect collapse in Fig.~\ref{fig:ising}(d) supporting the Gaussian behaviour at short times.

{\bf Conclusions:} The short-time dynamics of the IDLG, RDLG, and LG after a critical quench has been studied in two dimensions. We provide strong numerical evidence that transverse fluctuations in all these models at short times are governed by a non-interacting effective theory. 
Presumably, the conservation law renders the dynamics so slow that, before the model-specific correlations of the stationary state are built up, a long lapse emerges in which the behaviour is effectively Gaussian \footnote{The idea that the slow conservative dynamics is ultimately responsible for the observed Gaussian behaviour at short times is reinforced by the fact
that $g$ behaves similarly even when starting from a phase separated configuration, with the interface lying orthogonal to the drive (data not shown).}.
This agrees with the findings of Ref.~\cite{Tauber} concerning the evolution of $m \sim t^{0.5}$ and with those of Ref.~\cite{AS2002}, which showed that $O$ in IDLG and RDLG 
behave identically at short times. However, our analysis clarifies that this does not imply that these models belong to the same universality class.
Our results demonstrate, in fact, that it is not possible to distinguish the critical behaviour of IDLG from that of RDLG and hence to discriminate between the competing field theories, from the short-time dynamics of transverse observables alone. However, the long-time behaviour of the Binder cumulant shows that the two aforementioned models belong to two different universality classes. 

{\it Acknowledgements:} UB acknowledges the financial support by the ERC under Starting Grant 279391 EDEQS.

\newpage


\renewcommand{\thesection}{S\arabic{section}}
\renewcommand{\thetable}{S\arabic{table}}
\renewcommand{\thefigure}{S\arabic{figure}}
\renewcommand{\theequation}{S\arabic{equation}}

\begin{center}
 {\bf \large Supplemental Material for ``Short-time Behaviour  and Criticality of Driven Lattice Gases''}
\end{center}

\section*{Dynamical Behaviour of $O$ in Lattice Gases}

The dynamical behaviour of the anisotropic order parameter  $m$ [see Eq.~\eqref{eq:def-m} in the Letter] following a quench to the critical point is well described by
the Gaussian theory for all the three lattice gas models studied, $i.e.,$ driven lattice gas with either constant (IDLG) or random (RDLG) infinite drive and equilibrium lattice gas (LG). In other words, in the short-time regime, $m \sim t^{1/2}$ [see Eq. \eqref{eq:mt}] and the Binder cumulant $g$ of the lowest transverse mode [defined in Eq. \eqref{eq:binder}] is zero in this regime. The alternative order parameter $O,$ however, distinguishes between the driven (IDLG, RDLG) and the equilibrium (LG) lattice gases. 

In order to understand  this, we first write the phenomenological scaling form for $O$,  analogous to Eq. \eqref{eq:scalingass} in  the Letter,
\bea
O (t, L_{\parallel} ; S_\Delta) = L_{\parallel}^{-\beta/[\nu(1+\Delta)]} \tilde f_O (t/L_{\parallel}^{z/(1+\Delta)} ; S_\Delta).\quad
\label{eq:Oscalingass}
\eea
We already remarked that, in the LG, this scaling form is not compatible with the prediction $O \sim t^{1/8}  L_{\parallel}^{-1/2}$ of the Gaussian theory.  However, following Ref. \cite{AS2002}, it can be argued that, at short times, the only dependence of $O$ on the system size $L_{\parallel}$ is of the form $O \sim L_\parallel^{-1/2}$ which is very well confirmed by numerical simulations. Accordingly,  the generic behaviour of $O$ can be assumed to be
%
\bea
O \sim t^{\alpha} L_\parallel^{-1/2}, \label{eq:O}
\eea
where $\alpha$ is a phenomenological exponent to be determined. This, along with Eq. \eqref{eq:Oscalingass}, implies $\tilde f_O(x) \sim x^{\alpha}.$ Comparing the finite-size behaviour in Eq.~\eqref{eq:O} with Eq.~\eqref{eq:Oscalingass} one actually infers,
\bea
\alpha &=& \frac{1+ \Delta -2 \beta/\nu}{2 \, (4- \eta)}. \label{eq:alpha}
\eea
This equation, together with the hyperscaling relation $\Delta - 2 \beta/\nu= - \eta$ in two spatial dimensions, shows that the prediction $\alpha = 1/8$ of the Gaussian theory [see Eq. \eqref{eq:Ot}] can be obtained only when $\eta=0,$ which is the case for the IDLG (exactly) and the RDLG (approximately) but not for the LG. 

On the other hand,  Eq.~\eqref{eq:alpha} predicts $\alpha = 1/10$ upon substituting the values of the critical exponents corresponding to the Ising  universality class (LG). This is consistent with the numerical simulation results presented in the main text, see Fig. \ref{fig:ising}(b) therein.

\begin{figure}[th]
\vspace*{0.2 cm}
 \centering
 \includegraphics[width=10 cm]{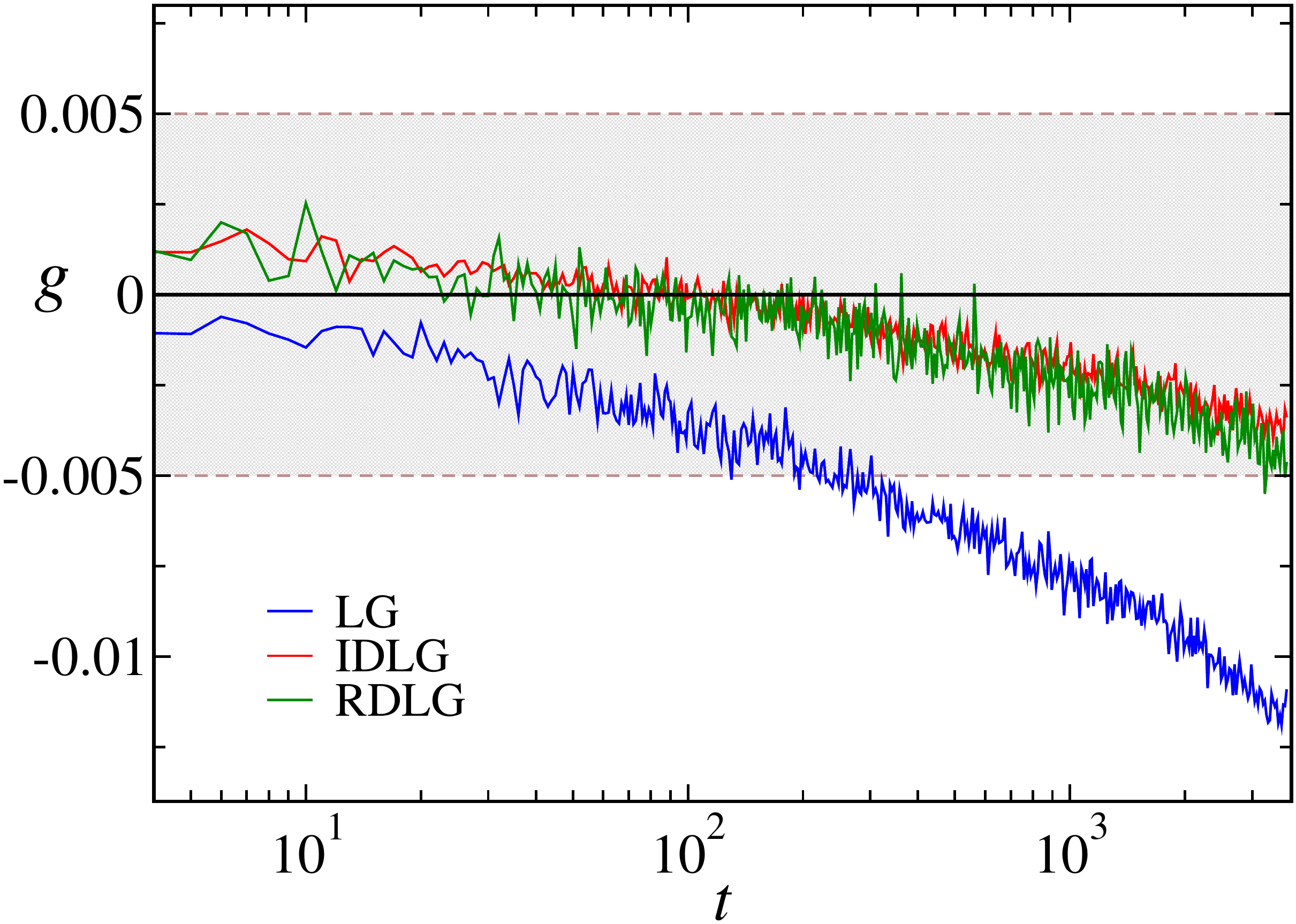}
\caption{Comparison between the temporal evolution of the Binder cumulants $g$ corresponding to the $12^{th}$ transverse mode, $i.e.,$ with $n_\perp =12,$ in the LG (lowest curve), IDLG and RDLG (two upper curves) on a $32 \times 32$ lattice. \label{fig:b}}
 \label{fig:binder}
\end{figure}

The emergence of this new value $1/10$ of the exponent $\alpha$ must be traced back to the non-Gaussian nature of higher fluctuating modes in the LG. In fact, even though the lowest mode behaves identically in all the three models we considered,  characterized by the same behaviour of $m$, higher modes show a significant difference in the non-driven case.

To illustrate this, we measured the Binder cumulants of higher modes which is defined  analogously to Eq.~(11), using transverse modes other than the first, i.e., with $\mu=\tilde \sigma(0,2 \pi n_\bot/L_\bot)$ and $n_\bot>1.$  
 Figure \ref{fig:b} compares the same for all the three lattice gases for the mode with $n_\perp =12$ on a $32 \times 32$ lattice. Clearly, the curve corresponding to the LG (lowest, blue) departs from Gaussian behaviour $g=0$ (in practice, $e.g.,$ $|g| \lesssim 0.005,$ corresponding to the shaded gray area) much earlier than it does for the IDLG  or RDLG (two upper curves, red and green respectively).

Accordingly, the different dynamical behaviour of $O$, which involves a sum over all modes, can be attributed to the non-Gaussian nature of the higher modes in the LG. 
Such a departure is not entirely surprising. In fact, for higher modes, mesoscopic descriptions such as the ones in Eqs. \eqref{eq:L-DLG} or \eqref{eq:g_evol} are not expected to hold, while the anisotropy at the microscopic level could be the mechanism leading to the Gaussianity of higher modes in the driven models.

 \end{document}